\documentclass{elsart}
\usepackage{epsf}
\begin{document}
\begin{frontmatter}
\title{Pairing-induced localization of the particle \\ 
          continuum in weakly bound nuclei}
\author[moscow]{S.A.~Fayans\thanksref{ELM}\thanksref{DFG}},
\author[moscow]{S.V.~Tolokonnikov\thanksref{DFG}},
\author[hannover]{D.~Zawischa}
\address[moscow]{Russian Research Centre -- Kurchatov Institute,
123182 Moscow, Russian Federation}
\address[hannover]{Institut f\"ur Theoretische Physik, 
Universit\"at Hannover, D-30060 Hannover, Germany}
\thanks[ELM]{E-mail address: fayans@mbslab.kiae.ru}
\thanks[DFG]{Supported by the Deutsche Forschungsgemeinschaft}
\begin{abstract}

The Hartree-Fock-Bogolyubov (HFB) problem for the cutoff local 
energy-density functional is solved numerically by using the 
Gor'kov formalism with an exact treatment of the 
particle continuum. The contributions from the resonant and ``gas''
continuum to the spectral density of the HFB eigenstates 
as well as the shifting and broadening of the discrete HF hole 
orbitals are clearly demonstrated with the illustrative example 
of the drip-line nucleus $^{70}$Ca. The structure of the neutron 
density distribution in the localized ground state is analyzed,
and the formation of its extended tail (``halo'') is shown to be 
a collective pairing effect. 



\end{abstract}
\end{frontmatter}

Pairing correlations significantly influence the properties of 
finite nuclei producing, e.g., the observed odd-even staggering 
of the binding energies and radii. In the mean-field models, the
pairing is introduced either at the BCS or the HFB 
level~\cite{RiSc80}. The zone of the active phase space around
the Fermi surface, in which the major pairing effects are developed, 
depends on the effective interaction used. Whenever this zone
includes states from the particle continuum, and this is obviously
inevitable when approaching the drip line, the BCS approximation
breaks down, yielding unphysical nucleonic gas~\cite{DFT84}.
A correct treatment of the pairing problem with continuum is
provided by the HFB theory which, for the chemical potential
$\mu<0$, always gives a localized ground state~\cite{Bul80}.
In principle, an exact solution of the HFB equations should be 
found with the physical boundary conditions both for the bound
and scattering states (examples of such solutions are given
in~\cite{BSTF87,STF88}). However, in most applications 
of the HFB theory to nuclei, with different kinds of the 
energy-density functionals, the particle continuum is discretized
in a box (see, e.g., \cite{DNW96,Miz99} and references therein),
and it is still questionable
whether or not the particle level density can be well reproduced
with such a prescription, particularly near the threshold. 
Continuum effects have been studied very recently in~\cite{BDP99}
within the HFB method using a fixed, analytically soluble mean 
field potential and keeping self-consistency only in the pairing 
channel; again, the continuum states have been discretized in a 
spatial box. An attempt to take into account the resonant continuum
within the HF+BCS approach has been also made recently 
in~\cite{SLW97,SGL98}. One should notice that, for a
nucleus with even $N$ (we consider the neutron subsystem for
definiteness), all HFB solutions with energies
$E>\vert\mu\vert\approx S_{2\rm n}/2$ ($S_{2\rm n}$ is the
two-neutron separation energy) belong to the continuum,
and these solutions contain both localized and non-localized HFB
resonances. Their positions and widths, as well as the genetic
relation to the Hartree-Fock (HF) single-particle spectrum, and also
the contribution from the nonresonant continuum are worth to be
revealed with numerically ``exact'' HFB calculations. Of particular
interest is the situation on the drip line where the nuclei have
only one bound state --- the ground state, and the whole HFB
spectrum is continuous. Our study is motivated by the great activity
in physics of radioactive nuclear beams which have lead already,
e.g., to the discovery of halo structure in some loosely bound
nuclei. 

Our approach is based on the generalized variational principle
applied to the cutoff (local) energy-density
functional~\cite{FTTZ98,FZmbx,FTTZ00} and on the physically
transparent and mathematically elegant Green's function method.
The coordinate-space technique used to solve the Gor'kov 
equations exactly, for spherical systems with a local
pairing field, is described in detail
in~\cite{BSTF87,FTTZ98,FTTZ00}. Here we shall discuss the results
for Ca isotopes obtained with functional DF3 whose normal part is
parametrized in Ref.~\cite{BFKZ96}. In the anomalous (pairing) part
we employ the contact ``gradient'' pairing force~\cite{FZ96} with a
density-dependent form factor
\begin{equation}
{\mathcal F}^\xi=1.35\times C_0\left(f^\xi_{\rm ex}+
h^\xi x^{2/3}(\vec r)+f^\xi_\nabla r_0^2
\left({\vec \nabla}x(\vec r)\right)^2\right)\,,\label{Fxi}
\end{equation}
where $x=(\rho_{\rm n}+\rho_{\rm p})/2\rho_0$ is the isoscalar 
dimensionless density with $\rho_{\rm n(p)}$ the neutron (proton) 
density and $\rho_0$ the equilibrium density of one kind of 
particles in symmetric nuclear matter with the Fermi energy 
$\epsilon_{0\rm F}$; $C_0=2\epsilon_{0\rm F}/3\rho_0$ is 
the inverse density of states at $\epsilon=\epsilon_{0\rm F}$; 
$r^2_0=(3/8\pi\rho_0)^{2/3}$ (for DF3, $C_0=308.2$~MeV$\cdot$fm$^3$
and $r_0=1.147$~fm). The pairing strength parameters
$f^\xi_{\rm ex}=-1.79$, $h^\xi=1.36$ and $f^\xi_\nabla=2.0$ have
been deduced from the neutron separation energies and charge radii 
in the Pb isotope chain~\cite{FTTZ00}, both the staggering and kink 
being well reproduced. As shown in Ref.~\cite{FTTZ00}, 
the force~(\ref{Fxi}) with the scaling factor of 1.35 
allows to nicely reproduce the analogous differential observables 
in the Ca isotopes including the anomalous behavior of charge 
radii. All calculations are done with a fixed energy cutoff of 
40~MeV above the Fermi level in any given nucleus.      

The HFB problem with the above energy-density functional is solved 
iteratively by using the coordinate-space technique in the 
complex energy plane. The normal and anomalous densities at each
iteration are evaluated as the contour integrals of the generalized 
Green's function $\hat G$ which obey the Gor'kov equation
$(E-\hat{\mathcal H})\hat G(E)=\hat I$ with $\hat {\mathcal H}$ the 
HFB Hamiltonian. For spherical systems, after separating the 
spin-angular variables one gets the equation for the radial 
component $\hat g_{jl}$ of the generalized Green's function 
$\hat G$: 
\begin{equation}
\left( \begin{array}{cc}
 E-h_{jl}+\mu & -\Delta \\
 -\Delta &  E+h_{jl}-\mu
\end{array} \right) 
\left(\begin{array}{cc}
g^{11}_{jl}  &\; g^{12}_{jl}\\
g^{21}_{jl}  &\; g^{22}_{jl}
	       \end{array} \right)(r,r';E)=
                   \left( \begin{array}{cc}
 \delta(r-r') & 0 \\
 0 & \delta(r-r')
	       \end{array} \right) \,,
\label{RGF}
\end{equation}
where $h_{jl}=\textstyle{\frac {\hbar^2}{2m}}
(-\textstyle{\frac{d^2}{dr^2}+\frac{l(l+1)}{r^2}})+
U_{\rm c}(r)+U_{sl}(r)\langle\vec{\sigma}\vec{l}\rangle_{jl}$ 
is the Hartree-Fock (HF) Hamiltonian in the $jl$ channel with 
$U_{\rm c}$ and $U_{sl}$ the central and spin-orbit potentials, 
respectively; $\langle\vec{\sigma}\vec{l}\rangle_{jl}=
j(j+1)-l(l+1)-\textstyle{\frac 3 4}$; $E\geq 0$. The solution of 
this matrix equation is
constructed by using the four linearly independent solutions 
$(u_i, v_i), i$=1--4, which satisfy the homogeneous system 
obtained from~(\ref{RGF}) by replacing the right hand side by
zero~\cite{BSTF87,FTTZ00}. Two of them are regular at 
$r\rightarrow 0$ and the other two are regular at 
$r\rightarrow\infty$. The latter are chosen with the asymptotic
momenta $k_\pm=\sqrt{\mu\pm E}$ such that ${\rm Im}k_\pm\geq 0$
which provides, in particular, the correct asymptotics for the
particle scattering states at the ${\rm Im}E=0$ axis. Such
boundary conditions ensure the spatial localization of the nuclear
ground state at $\mu<0$~\cite{Bul80,BSTF87}.

It is convenient to study the pairing effects in terms of the 
spectral distributions over the eigenstates of the HFB Hamiltonian, 
i.e. in terms of the spatial integrals  of the imaginary parts of 
the Gor'kov Green's functions. Here, for the given $jl$ quantum
numbers, we shall consider two partial spectral distributions:
\begin{equation} 
\frac{\d N^{22}_{jl}}{\d E}=-\frac{2j+1}{\pi}
\int_0^R{\rm Im}g^{22}_{jl}(r,r;E)\d r\,,
\label{njlvac}
\end{equation}
and
\begin{equation} 
\frac{\d N^{11}_{jl}}{\d E}=-\frac{2j+1}{\pi}
\int_0^R{\rm Im}g^{11}_{jl}(r,r;E)\d r\,.
\label{njlcon}
\end{equation}
The first and the second we shall call the $v^2$- and $u^2$-spectral 
density, respectively. The corresponding analytical expressions  
can be obtained by using the partial Green's functions which 
incorporate the generalized Wronskian of the four linearly independent 
solutions as given in Refs.~\cite{BSTF87,FTTZ00}. We add that the 
expressions for the canonical-basis spectral distributions and the 
numerical results with discretized continuum for a few tin isotopes 
can be found in Ref.~\cite{DNW96}. 

In eq.~(\ref{njlvac}), the spatial integral converges for 
$R\rightarrow\infty$ since it includes only the localized components 
$\propto{v^2}$. At $E>\vert\mu\vert$, all HFB eigenstates are in the 
continuum and, since ${\rm Im}g^{11}_{jl}$ contains non-localized 
components $\propto{u^2}$, the integration in eq.~(\ref{njlcon}) is 
meaningful only up to a certain finite range $R$. One may present the 
two above expressions by one function $\d N_{jl}^{\rm HFB}/\d\epsilon$ 
if~(\ref{njlvac}) and~(\ref{njlcon}) are considered as functions of 
$\epsilon=\mu -E$ and $\epsilon=\mu +E$, respectively, though both are 
for the same HFB eigenvalue $E$.  
\begin{figure}[t] 
\begin{center} 
\epsfxsize=22pc 
\epsfbox{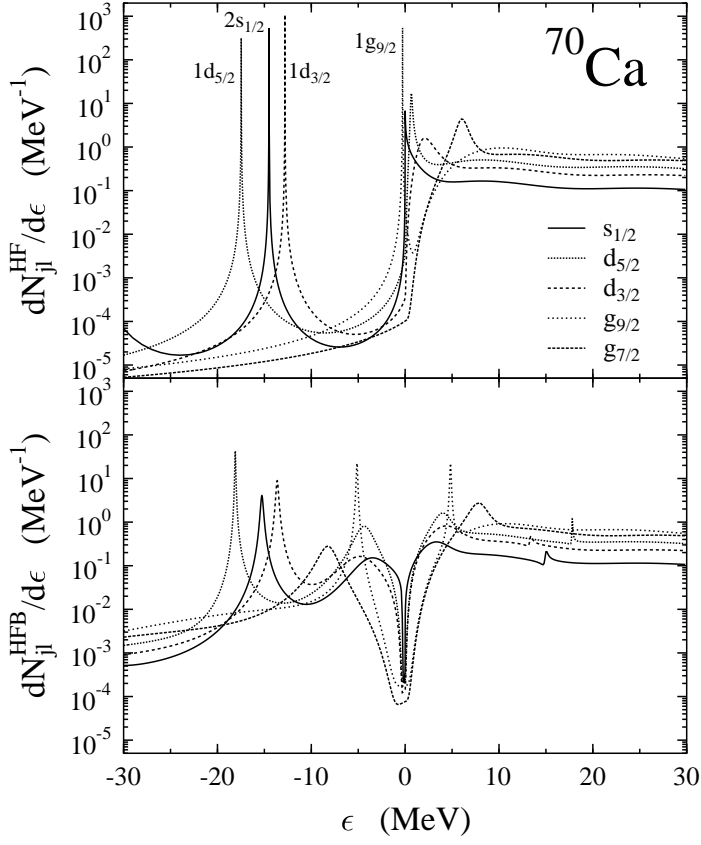} 
\end{center}
\caption{Neutron even-$l$ partial spectral densities in drip-line 
nucleus $^{70}$Ca calculated by integrating the imaginary parts of 
the corresponding Green's functions up to $R=10$~fm. A smearing 
parameter $\gamma =1$~keV is used to visualize the $\delta$-like 
peaks and very narrow resonances. Upper panel: level densities for
the ``HF'' potential; the $\epsilon<0$ part for the discrete
spectrum with indicated single-particle levels, the $\epsilon>0$
part for continuum states. Lower panel: the HFB results; all states
are in the continuum, the left part ($\epsilon<0$) is for the
localized $v^2$-spectral density, the right ($\epsilon>0$) for the
non-normalizable $u^2$-components (see text).} \label{f:f1} 
\end{figure}
\begin{figure}[t]
\begin{center}
\epsfxsize=22pc 
\epsfbox{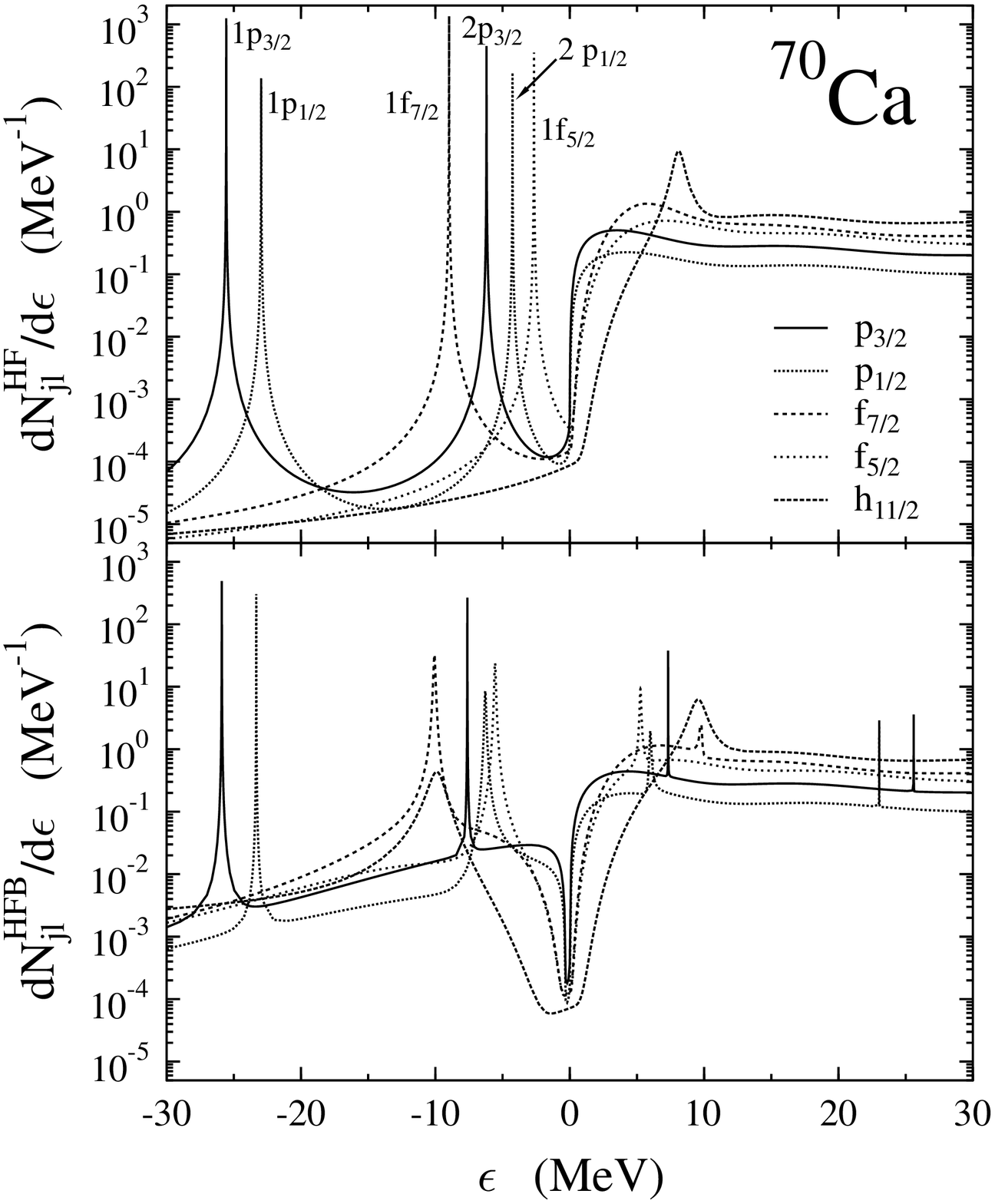} 
\end{center}
\caption{The same as in Fig.~1 but for the odd-$l$ states.} 
\label{f:f2} 
\end{figure}
We show in Figs.~1 and 2 the partial HFB neutron spectral densities 
for even and odd $l$, respectively, calculated for the drip-line 
nucleus $^{70}$Ca, as functions of $\epsilon$ with $R$=10~fm (lower 
panels) and also present the corresponding ``HF'' level densities 
(upper panels). The ``HF'' refers to the results obtained by solving
eq.~(\ref{RGF}) but keeping only the diagonal part of 
$\hat {\mathcal H}$ which emerges after solving the HFB 
problem ($h_{jl}$ includes the contribution from pairing correlations). 
In other words, in these figures we show the level densities for the 
basis which diagonalizes the mean field part of the HFB Hamiltonian 
versus the spectral densities which arise when the pairing field 
$\Delta$ is present explicitly in the left matrix in 
eq.~(\ref{RGF}). We point out that the spherical volume of radius 
$R$=10~fm is chosen here as the physical volume in which the 
mean-field and pairing correlations are mostly active to produce the 
localized ground state, it is not a normalization box; within this 
volume we integrate the imaginary parts of the Green's functions which 
have the
correct asymptotic behavior (the normal and anomalous densities, when 
solving the HFB problem, have been evaluated up to 25~fm, and, again, 
the wave functions are not zero at and beyond this radius, but the 
contribution from this far region to the bulk ground state properties 
is negligible). For the visualization of the $\delta$-like peaks 
associated with the discrete ``HF'' states and very narrow resonances, 
a small imaginary part $\gamma$=1~keV has been added to the energy 
$E$ in the Green's functions. The obtained chemical potential is 
small, $\mu=-149$~keV; the spectral densities within the region 
$2\mu<\epsilon<0$ should vanish, and those which appear in 
Figs.~1 and 2 at these energies are due to the smearing parameter 
$\gamma$.    

Let us discuss first the peculiarities in the ``HF'' level densities. 
For a finite-depth potential well, such as our neutron HF potential, 
it can be shown that the partial single particle level density in 
the continuum within a spherical volume with radius $R$ larger than 
the potential range is given by 
\begin{equation}
\frac{\d N^{\rm HF}_{jl}}{\d\epsilon}=\frac{2j+1}{\pi}
\left\{\frac{\d\delta_{jl}}{\d\epsilon}+\frac{k^3R^3}{2\epsilon}
\left[\left(Y^l_{jl}(kR)\right)^2
-Y^{l-1}_{jl}(kR)Y^{l+1}_{jl}(kR)\right]\right\}\,,\label{levhf}
\end{equation}
where $Y^i_{jl}(kR)=(\cos\delta_{jl})j_i(kR)-
(\sin\delta_{jl})n_i(kR),\,i=l,l\pm 1;\,k=\sqrt{2m\epsilon}/\hbar$. 
This formula is valid for $R$ big enough so that the radial wave 
function can be expressed through spherical Bessel and Neumann 
functions ($j_l$ and $n_l$, respectively) and the phase shift 
$\delta_{jl}(\epsilon)$ can be calculated accurately. The 
resonance-like bumps seen in the upper panels in Figs.~1 and 2 at 
$\epsilon>0$ are associated with the term 
$\d\delta_{jl}/\!\d\epsilon$ in~(\ref{levhf}). These bumps are 
imposed on a rather flat nonresonant continuum which weakly 
oscillates with an energy-dependent wavelength defined by 
$\Delta(kR)\approx 2\pi$. Without potential, when 
$\delta_{jl}(\epsilon)\equiv 0$, eq.~(\ref{levhf}) gives the 
free-gas partial level density. Subtracting the latter from 
eq.~(\ref{levhf}), at $kR\gg l$ one gets
\begin{equation}
\frac{\d N^{\rm HF}_{jl}}{\d\epsilon}-
\frac{\d N^{\rm gas}_{jl}}{\d\epsilon}
=\frac{2j+1}{\pi}
\left[\frac{\d\delta_{jl}}{\d\epsilon}
-\cos(2kR+\delta_{lj})\frac{\sin\delta_{lj}}{2\epsilon}\right]\,.
\label{krl}
\end{equation}
Averaging this expression over small energy range 
$\Delta\epsilon>\pi\sqrt{2\hbar^2\epsilon/mR^2}$ 
eliminates the oscillating cosine term, and one is left with the
usual level density $\propto\d\delta_{jl}/\!\d\epsilon$ associated 
with the potential well itself (see also Ref.~\cite{SKD97}).
However, this procedure can not be applied at low energies, 
i.e. in the region near the continuum threshold which, in drip-line 
nuclei, is very close to the Fermi surface. This zone of the phase 
space is of importance since the major pairing effects are developed
around the Fermi surface. At any finite $R$, when 
$\epsilon\rightarrow 0$, the leading term $\propto k^{2l-1}$
of the expansion of $\d\delta_{jl}/\!\d\epsilon$ in powers of $k$ 
cancels in eq.~(\ref{levhf}) due to the contribution coming from
the term $\propto\sin(2\delta_{jl})j_{l-1}(kR)n_{l+1}(kR)$. The
resulting level density near the continuum threshold turns out to be 
proportional to $C_{jl}k^{2l+1}$, i.e. it has the free-gas behavior, 
the coefficients $C_{jl}$ depend, however, on the potential well.
For the $s$-wave level density, using the effective range 
approximation, we get from~(\ref{levhf}) and~(\ref{krl})
an analytical expression: 
\begin{equation}
\frac{\d N^{\rm HF}_{\frac{1}{2}0}}{\d\epsilon}=\frac{2}{3\pi}
\frac{k^3}{\epsilon}R^3
\left\{1-3\frac{a}{R}\left[1-\frac{a}{R}\left(1-\frac{1}{2}
\frac{r_{\rm eff}}{R}\right)\right]\right\}\,.\label{effr}
\end{equation}
where $a$ is the scattering length and $r_{\rm eff}$ is the 
effective range. In the curly brackets here is a factor which could
give an enhancement of the HF level density as compared to the 
free-gas one. For our ``HF'' potential we found $a=-14.8$~fm and 
$r_{\rm eff}=14.4$~fm, which gives, with $R=10$~fm, an enhancement 
factor of about 7. These parameters correspond to the virtual 
$s$-state at $\epsilon_0=-51$~keV which shows up as a sharp peak at 
very low positive $\epsilon$ in the upper panel in Fig.~1. The
width of this very asymmetric peak at half maximum is about
23~keV. At larger energies, one can see in this panel two partners
of the $d$-wave resonance which is split due to the spin-orbit
potential, and also the $g_{7/2}$ resonance whose $g_{9/2}$ partner
is in the discrete spectrum at $\epsilon =-276$~keV. Besides, a 
negative-parity $h_{11/2}$ resonance can be seen in the upper panel 
in Fig.~2. The resonance characteristics and the energies 
$\epsilon_\lambda^0$ of some bound ``HF'' states are listed in 
Table~1 (the first three columns). 

\begin{table}[b]
\begin{description}
\item[Table 1.]
\begin{center}
Characteristics$\,^a$ of some ``HF'' states and localized\\
$v^2$-components of the HFB resonances in $^{70}$Ca \\
\end{center}
\end{description}
$$\mbox{%
\begin{tabular}{llclll} 
\hline
\multicolumn{1}{c}{state}  &
\multicolumn{2}{c}{``HF''} & &
\multicolumn{2}{c}{HFB}\\ 
\cline{2-3} \cline{5-6}
\multicolumn{1}{c}{$\lambda$} &
\multicolumn{1}{c}{$\epsilon^0_\lambda$} &
\multicolumn{1}{c}{$\Gamma^0_\lambda$}   & & 
\multicolumn{1}{c}{$\epsilon_\lambda$} &
\multicolumn{1}{c}{$\Gamma_\lambda$}\\ 
\hline
3s$_{1/2}$ &$-$0.051$\,^b$&$\approx0.023\,^b$&&$-$3.467&4.574\\
2d$_{5/2}$ &   +0.659  &  0.192 && $-$4.369 &  2.150  \\
2d$_{3/2}$ &   +2.097  &  2.038 && $-$4.859 &  4.483  \\
1g$_{9/2}$ & $-$0.276  &        && $-$5.132 &  0.144  \\
1f$_{5/2}$ & $-$2.666  &        && $-$5.559 &  0.115  \\
2p$_{1/2}$ & $-$4.261  &        && $-$6.285 &  0.122  \\
2p$_{3/2}$ & $-$6.193  &        && $-$7.621 &$5\times10^{-5}$\\
1g$_{7/2}$ &   +6.064  &  1.419 && $-$8.236 &  2.115  \\
1h$_{11/2}$&   +8.106  &  0.915 && $-$9.896 &  1.330  \\
1f$_{7/2}$ & $-$8.980  &        &&$-$10.071 &  0.145  \\
1d$_{3/2}$ &$-$12.814  &        &&$-$13.663 &  0.277  \\
2s$_{1/2}$ &$-$14.514  &        &&$-$15.265 &  0.306  \\
1d$_{5/2}$ &$-$17.467  &        &&$-$18.103 &  0.086  \\
1p$_{1/2}$ &$-$22.962  &        &&$-$23.328 &$1\times10^{-4}$\\
1p$_{3/2}$ &$-$25.564  &        &&$-$25.894 &  0.003  \\
\hline
\multicolumn{6}{l}{$^a$ Energies and widths are in MeV.}\\ 
\multicolumn{6}{l}{$^b$ Virtual state (see text).}\\ 
\end{tabular}}$$
\end{table}
The graphical presentation of Fig.~1 and~2 has the advantage that 
the HFB spectral densities from the lower panels would be smoothly
transformed to the corresponding upper ``HF'' ones if the
$\Delta$'s in the HFB Hamiltonian of eq.~(\ref{RGF})
could be smoothly switched off keeping the ``HF'' part $h_{jl}$ 
unchanged and defining $\mu$ by the particle number condition. 
In this way, one can easily trace the genetic origin 
of the irregularities seen in the HFB spectral densities.
In particular, the resonance contributions from the particle 
continuum to the localized $v^2$-spectral density can be recognized.
The characteristics of the picked-out localized $v^2$-parts of the
HFB resonances are listed in Table~1 attributing them the genetic
``HF'' $jl$ quantum numbers (the fourth and fifth columns). The
widths of the ``HF'' resonance states and of the localized
$v^2$-components of the HFB resonances have been estimated
from a 3-point no-background Breit-Wigner fit of the corresponding
maxima. One can see that the discrete ``HF'' states get widths, i.e.
they become the HFB resonances whose $v^2$-components are shifted,
as expected, to lower $\epsilon$. The HFB nonresonant background can
be also seen in the lower panels of Figs.~1 and 2. The non-localized
$u^2$- and localized $v^2$-components of the HFB resonances are
located symmetrically with respect to $\epsilon =\mu$ (i.e., $E=0$).
We remark that all these belong to the same HFB eigenstate, however
we do not discuss here the properties of the $u^2$-components due to
lack of space. 

As seen in Fig.~1, the $1g_{9/2}$ ``HF'' state at 
$\epsilon_\lambda^0=-276$~keV becomes, with pairing, a HFB resonance
whose localized $v^2$-component lies well below the continuum 
threshold, at $\epsilon\approx -\bar{\Delta}$ where $\bar{\Delta}$
is the diagonal matrix element of the pairing field over the ``HF''
states near the Fermi surface. In our case, the energy gap
$2\bar\Delta$ between the two components of this resonance is rather
big, $\approx 10$~MeV. This could lead to the noticeable reduction
of the tail of the neutron density distribution (the so-called
pairing anti-halo effect~\cite{BDP99,Miz99}). However, in the 
drip-line nuclei, when $\vert\mu\vert\ll\bar\Delta$, near the
threshold there appear strong localized $v^2$-components from the
resonant and non-resonant parts of the particle continuum with
different $jl$. This is clearly seen in the lower panels of
Figs.~1 and~2. 
\begin{figure}[t]
\begin{center}
\epsfxsize=22pc 
\epsfbox{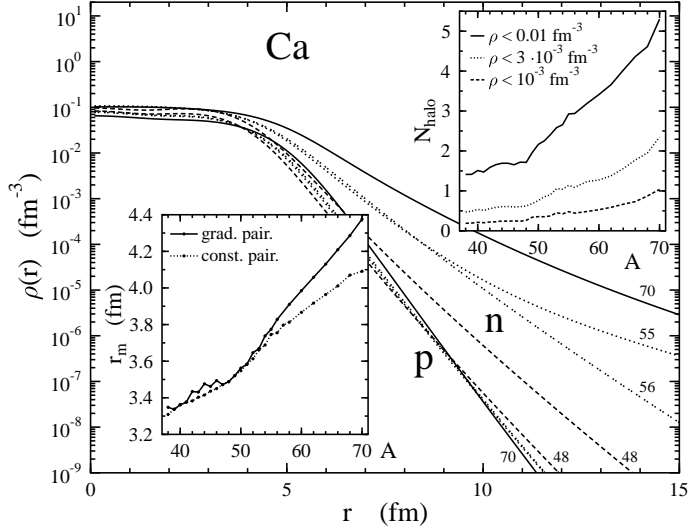} 
\end{center}
\caption{Neutron and proton densities for some Ca isotopes
resulting from the spherical HFB calculations with functional
DF3 and gradient pairing force of eq.(\ref{Fxi}). The curves are 
marked with the mass number $A$. The two inserts are both for even-
and odd-$A$, up to the drip-line nuclei $^{70}$Ca and $^{55}$Ca,
respectively. Right upper insert: the number of neutrons in the
``halo'' region where the density does not exceed the indicated
values (0.01, 0.003 and 0.001 fm$^{-3}$). Left lower insert: the rms
matter radii in comparison with ``constant-pairing'' HFB results
(see text).} \label{f:f3} 
\end{figure}

\begin{figure}[t]
\begin{center}
\epsfxsize=22pc 
\epsfbox{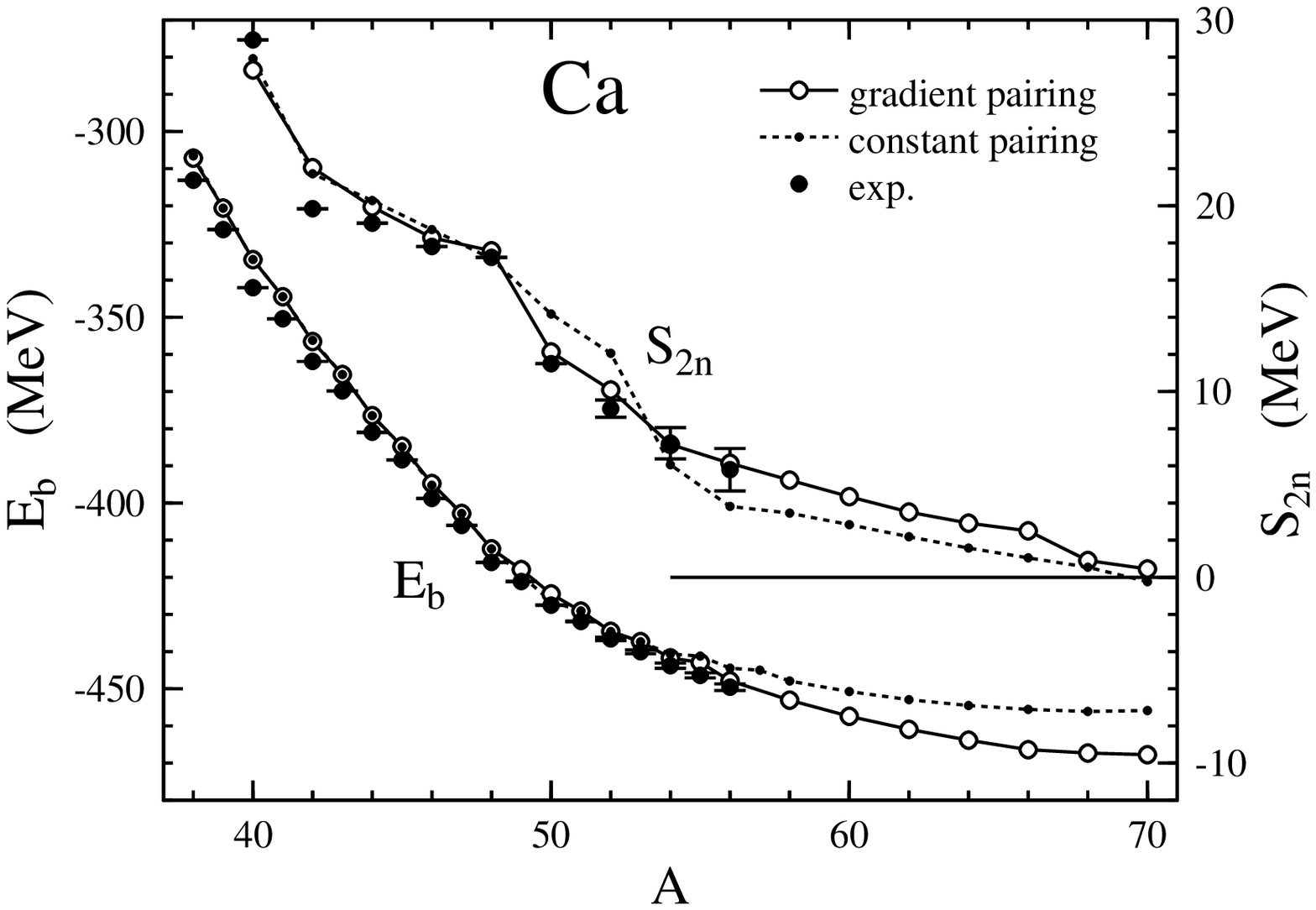} 
\end{center}
\caption{Binding energies $E_{\rm b}$ (lower curves, left scale) 
and two-neutron separation energies $S_{2{\rm n}}$ (upper curves, 
right scale) in even Ca isotopes. Solid (dashed) lines connect 
the points resulting from the spherical HFB calculations with 
functional DF3 and gradient (``constant'') pairing force (see 
text). Experimental data and those derived from systematics are 
from~\cite{AW93}.} \label{f:f4} 
\end{figure}

The ``HF'' and HFB partial neutron occupation numbers $N_{jl}$ in
$^{70}$Ca are listed in Table~2. Also given in this table are the
numbers of neutrons $N_{jl}(r>R_1)$ and $N_{jl}(r>R_2)$ in the tail
of $\rho_{\rm n}$ beyond the nuclear surface; $R_1$ and $R_2$ are
defined by $\rho_{\rm n}(R_1)=0.01$~fm$^{-3}$ and 
$\rho_{\rm n}(R_2)=0.001$~fm$^{-3}$, respectively. The choice of
$R_1$ roughly corresponds to the radius $R_{\rm F}+t/2$ where
$t=4(\ln3)d$ is the width of the diffuse zone approximated in the
usual way by the Fermi function 
$\propto \{1+\exp[(r-R_{\rm F})/d]\}^{-1}$. The calculated densities
for Ca isotopes with $A=38$ to 70 (a few of them are shown in
Fig.~3) can be well described in the surface region by such a Fermi
distribution. For $^{70}$Ca, we found $R_{\rm F}=4.60$~fm and
$d=0.73$~fm which gives $R_1=6.2$~fm (for comparison: in $^{48}$Ca,
$R_{\rm F}=3.85$~fm, $d=0.52$~fm and $R_1=5.0$~fm; thus, approaching
the drip line, the size of the diffuse surface zone is significantly
increasing). Using a larger radius $R_2$, one can characterize the
$jl$ composition of the outmost tail of $\rho_{\rm n}$ in the
``halo'' region. Table~2 shows that the pairing makes the $N_{jl}$
distribution more ``uniform'' than in the ``HF'' case, the major
effect is due to the depletion of the discrete ``HF'' orbitals
$g_{9/2}$, $f_{7/2}$ and $f_{5/2}$ and filling the resonance
structures from the ``HF'' continuum ($3s_{1/2}$, $2d_{5/2}$, etc.,
see Table~1). The partial neutron numbers in the ``halo'' region are
also distributed more uniformly, the contributions from the states
with higher angular momenta, up to $l=8$, are not negligible. Thus,
in our calculations, the composition of the extended tail of the
neutron density distribution in $^{70}$Ca is determined by a
collective pairing effect.
\begin{table}[t]
\begin{description}
\item[Table 2.]
\begin{center}
Partial neutron occupation numbers in $^{70}$Ca \\
\end{center}
\end{description}
$$\mbox{%
\begin{tabular}{lcccccccc}
\hline
 & \multicolumn{2}{c}{$N_{jl}$}          & &
   \multicolumn{2}{c}{$N_{jl}(r>R_1)\,^a$} & & 
   \multicolumn{2}{c}{$N_{jl}(r>R_2)\,^b$} \\
\cline{2-3}  \cline{5-6}  \cline{8-9}
\multicolumn{1}{c}{$lj$}    &
\multicolumn{1}{c}{``HF''}  &
\multicolumn{1}{c}{HFB}     & &
\multicolumn{1}{c}{``HF''}  &
\multicolumn{1}{c}{HFB}     & &
\multicolumn{1}{c}{``HF''}  &
\multicolumn{1}{c}{HFB}\\
& & & &\multicolumn{1}{c}{$R_1$=6.10~fm} &
\multicolumn{1}{c}{$R_1$=6.15~fm}        & &
\multicolumn{1}{c}{$R_2$=7.60~fm}        &
\multicolumn{1}{c}{$R_2$=8.15~fm}\\
\hline
s$_{1/2}$ & 4&4.740&&0.079&0.501&&0.007&0.176 \\
p$_{3/2}$ & 8&7.861&&0.612&0.711&&0.120&0.146 \\
p$_{1/2}$ & 4&3.774&&0.408&0.393&&0.104&0.083 \\
d$_{5/2}$ & 6&8.207&&0.094&0.920&&0.006&0.204 \\
d$_{3/2}$ & 4&4.725&&0.088&0.439&&0.008&0.103 \\
f$_{7/2}$ & 8&8.012&&0.362&0.554&&0.039&0.077 \\
f$_{5/2}$ & 6&4.599&&0.589&0.503&&0.129&0.073 \\
g$_{9/2}$ &10&5.274&&1.267&0.633&&0.294&0.072 \\
g$_{7/2}$ &  &0.877&&     &0.211&&     &0.035 \\
h$_{11/2}$&  &0.925&&     &0.191&&     &0.026 \\
h$_{9/2}$ &  &0.292&&     &0.097&&     &0.018 \\
i$_{13/2}$&  &0.321&&     &0.087&&     &0.014 \\
i$_{11/2}$&  &0.136&&     &0.059&&     &0.011 \\
j$_{15/2}$&  &0.144&&     &0.052&&     &0.009 \\
j$_{13/2}$&  &0.054&&     &0.035&&     &0.008 \\
k$_{17/2}$&  &0.040&&     &0.026&&     &0.006 \\
k$_{15/2}$&  &0.019&&     &0.016&&     &0.005 \\
\hline
$\Sigma_{lj}\,$:&50&50.00&&3.500&5.429&&0.707&1.067\\
\hline
\multicolumn{9}{l}{$^a$ $R_1$ is defined by the condition
$\rho_{\rm n}(R_1)=0.01$~fm$^{-3}$} \\
\multicolumn{9}{l}{$^b$ $R_2$ is defined by the condition
$\rho_{\rm n}(R_2)=0.001$~fm$^{-3}$} \\
\end{tabular}}$$
\end{table}

The numbers of neutrons beyond the diffusive surface,
$N_{\rm halo}$, calculated both for even and odd Ca isotopes with
$A=38$ to 70 are shown in the right insert in Fig.~3. Depending on
the choice of the upper limit for the tail density, 0.01, 0.003 or
0.001 fm$^{-3}$, $N_{\rm halo}$ varies from 1.41 to 5.30, from 0.47
to 2.35 and from 0.19 to 1.04, respectively; the relative increase
of $N_{\rm halo}$ for the latter choice reaches a factor of 5.5.
Irregulaties seen in in the behavior of $N_{\rm halo}$ are due to
the odd-even effect. Odd isotopes are calculated with
blocking~\cite{FTTZ00}, the heaviest one is predicted to be
$^{55}$Ca. The neutron density in this drip-line nucleus is also
plotted in Fig.~3, the one-neutron ``halo'' tail of $\rho_{\rm n}$
in this case originates from the $2p_{1/2}$ blocked level with
$\epsilon_\lambda =-820$~keV. 

The rms matter radii $r_{\rm m}$ are shown in the left insert in
Fig.~3. The staggering seen in the lighter Ca isotopes is generated
by the gradient pairing force~(\ref{Fxi}) whose parameters were
chosen to reproduce both the neutron separation energies $S_{\rm n}$
and anomalous behavior of charge radii~\cite{FTTZ00}. One may
suspect that this force yields too strong pairing near the drip line
where the surface becomes more diffuse and the repulsive gradient
term in~(\ref{Fxi}) decreases; as the result we get 
$\bar\Delta\approx 5$~MeV for $^{70}$Ca and $\bar\Delta\simeq
1.8$~MeV around $^{44}$Ca, the latter being in agreement with
experiment. Unfortunately, the density dependence of the effective
pairing force is not well known, and the extrapolation to the
drip line with parametrization~(\ref{Fxi}) may be questionable.
Quite different results could be obtained with another kind of the
pairing force. One can use, for example, instead of~(\ref{Fxi}),
a simple $\delta$-interaction without $\rho$-dependence, with
${\mathcal F}^\xi=-0.66C_0$, which also reproduces well
the experimental $S_{\rm n}$ values, and calculate then the whole Ca
chain. The matter radii obtained with such a ``constant'' force are
shown by the dotted line in the left insert in Fig.~3, and the
binding energies together with two-neutron separation energies are
presented in Fig.~4 in comparison with the results for the
force~(\ref{Fxi}). The behavior of $r_{\rm m}$ versus $A$ for
``constant'' ${\mathcal F}^\xi$ is relatively smooth, without
noticeable staggering~\footnote{We remark that neither the observed 
evolution of charge radii in Ca nor the kink in Pb isotopes could be 
reproduced with this simple pairing force~\cite{FTTZ00} which is thus 
rather unrealistic and serves here only as a demonstrative example.}; the 
odd- and even-$A$ drip-line nuclei occur to be $^{57}$Ca and $^{70}$Ca, 
respectively. Around these nuclei, weak irregularities can be seen in the 
behavior of $r_{\rm m}$. $^{70}$Ca comes out with $\Delta=0$, i.e. pairing 
correlations do not appear for ``constant'' ${\mathcal F}^\xi$ in
this nucleus; besides, the calculated $S_{2\rm n}$ value is
negative, $-236$~keV (see Fig.~4), so $^{70}$Ca could be a
two-neutron drip-line double-magic emitter. As seen in Fig.~4, the
binding energies $E_{\rm b}$ and $S_{2\rm n}$ values are reproduced
fairly well with both kinds of the pairing force, though a better
description is obtained with the gradient force~(\ref{Fxi}).
Systematic deviations start at the mass numbers right beyond the
measured region. Clearly, new experimental data and more extensive
theoretical studies are needed to find still better parametrizations
of the effective pairing force and improve the predictive power
of self-consistent mean-field models.    

In conclusion: we have presented a numerically exact solution of
the HFB problem for the cutoff local energy-density functional. The
Green's function formalism with physical boundary conditions has
been used. The localization of the particle continuum in the drip
line nucleus $^{70}$Ca has been analyzed in terms of the partial
spectral densities for the HFB eigenstates whose spectrum is fully
continuous. The genetic relation to the ``HF'' (mean-field) partial 
level densities has been discussed, and the contributions from the
resonant and nonresonant particle continuum to the localized ground
state, as well as the shifting and broadening of the discrete hole
orbitals have been clearly revealed. The composition of the
extended neutron distribution in the ``halo'' region beyond the
diffuse surface has beeen analyzed in terms of the partial occupation
numbers, and its formation has been shown to be a collective pairing
effect.    

\ack{Partial support of this work by the Deutsche
Forschungsgemeinschaft and by the Russian Foundation for Basic
Research (project 98-02-16979) is gratefully acknowledged.}

\end{document}